# MI-PEFT: Mixture-of-Experts Integrated Parameter-Efficient Fine-Tuning Protein Language Models Improves Acidophilic Proteins Classification

Honghan Shen

University of Wisconsin–Madison, Madison, WI 53706, USA,

hshen87@wisc.edu

**Abstract.** Acidophilic proteins that remain stable and functional under highly acidic conditions, are important for industrial biocatalysis, acid-related bioprocessing, and the discovery of acid-stable enzymes. However, their identification relies heavily on time-consuming experimental screening methods. With the rapid growth of protein sequence databases, the need for computational identification methods that are both accurate and efficient has become stronger. The emergence of protein language models (PLMs) has significantly improved the sequence representation of downstream biological prediction tasks. This paper proposes MI-PEFT, a mixture-of-experts integrated parameter-efficient fine-tuning framework. Built on the ESM C-600M backbone, the framework incorporates LoRA-based PEFT methods and a DeepSeekMoE-based classification head to resolve the limitations of PEFT and significantly improve computational efficiency. Notably, this task is characterized by a significant class imbalance in the dataset, making high specificity particularly challenging. The experimental results demonstrate that MI-PEFT on PLMs, especially $C^3A$, serves as an efficient tool for identifying acidophilic proteins and a constrained pathway that helps resolve class-imbalance by preserving the pretrained representations.

## 1 Introduction

Acidophilic proteins, which maintain their structural integrity and catalytic function under extremely low pH conditions (pH<4), are of immense practical value in industrial biocatalysis and acid-related bioprocessing (Sharma et al., 2012). Their stability in harsh environments makes them ideal for applications such as bioleaching, acid mine drainage treatment, and production of acid-stable enzymes (Lu & Liu, 2026, Mesbah et al., 2022, Kour et al., 2019). However, identification of these proteins has long relied on wet-lab experiments, which are labor-intensive, time-consuming, and costly (Song et al., 2025).

Earlier computational approaches mainly relied on handcrafted descriptors such as amino-acid composition, k-mer statistics, and PSSM-derived features coupled with shallow classifiers (Ofer et al., 2021; Lin et al., 2013). Although useful, these representations compress sequences into low-order summaries and may miss context-dependent signals relevant to acidophilic adaptation (Fang et al., 2012).

In this context, Protein Language Models (PLMs) trained on mass protein datasets provide richer sequence representations than handcrafted composition features (Asgari, et al., 2015; Rao et al., 2019; Elnaggar et al., 2022). Among PLM families, Bidirectional Encoder Representations from Transformers-style (BERT) masked language modeling has been particularly influential (Devlin et al., 2018) as it learns bidirectional context by predicting masked residues. In particular, the ESM family (Rives et al., 2021) demonstrated that scaling unsupervised transformer language modeling to hundreds of millions of protein sequences yields embeddings in which biologically meaningful information about structure and function emerges, enabling accurate inference even without handcrafted features (Zheng, et al., 2024).

Recent state-of-the-art work on in-silico acidophilic protein identification (Susanty, et al., 2024) has demonstrated that pretrained protein language model embeddings can serve as highly informative representations for downstream classification, enabling strong performance even when coupled with lightweight classifiers. The combination of Logistic Regression and embeddings reduces GPU memory overhead and makes the pipeline accessible for limited-resource settings.

To further adapt the model to a specific task while preserving computational efficiency, this study adopts a Parameter-Efficient Fine-Tuning (PEFT) framework that adapts ESM Cambrian (ESM C) backbones, updating only a small fraction of parameters and dramatically reducing computational and memory costs while maintaining competitive performance (Schmirler et al., 2024). Concretely, we fine-tuned ESM C with six representative LoRA-based PEFT methods, including LoRA (Hu et al., 2021) , MiSS (Kang et al., 2024), $C^3A$ (Chen et al., 2024), OFT (Qiu et al., 2023), RandLoRA (Albert et al., 2025), and SHiRA (Bhardwaj et al., 2024).These adapters update the key projection according to a specific task while keeping the number of trainable parameters extremely small. In addition, motivated by recent advances in sparse Mixture-of-Experts architectures, we replaced a linear classification head with a DeepSeekMoE-based expert router that activates only a small subset of experts per sample (Dai et al., 2024), which increases classifier expressiveness and significantly reduces trainable parameters. Finally, to mitigate potential miscalibration and overconfidence in neural predictions, we applied post-hoc temperature scaling to the validation split and evaluated the calibrated logits on the test fold, yielding more reliable probability estimates. Across the same benchmark dataset and evaluation protocol, we conducted systematic experiments where we evaluated the performance using 10-fold cross-validation on the training set, reported the final results on a held-out independent test set, and reported six standard metrics—Accuracy, Sensitivity, Specificity, Matthews Correlation Coefficient, F1-score, and ROC-AUC. The results demonstrated that all six approaches outperformed basic machine learning, deep learning methods, and the previous SOTA baseline and $C^3A$ achieved the best performance.

## 2 Methods

### 2.1 Dataset

The benchmark dataset used in this study adopts the established dataset in prior acidophilic protein prediction studies and was constructed from publicly available biological repositories (Sayers et al., 2021). Specifically, candidate sequences were retrieved from the National Center for Biotechnology Information (NCBI) databases, which provide comprehensive and continuously updated protein records and annotations. In a previous study, a large set of candidate acidophilic proteins was collected by searching for proteins annotated as acidophilic, which naturally produced far more positives (237,109) than negatives (7,394), reflecting both the availability of annotated acidophilic entries and the focus of existing sequence repositories (Susanty et al., 2024). For the non-acidophilic class, rather than treating all unannotated proteins as negatives, a contrast set of clearly characterized extremophile proteins (thermophilic, halophilic, and alkaliphilic) was assembled because proteins without an acidophilic label may still exhibit partial acid tolerance or belong to mixed categories such as thermo-acidophilic proteins. Entries containing ambiguous amino-acid symbols and fragments shorter than 100 residues were removed to ensure high-quality sequences. Redundancy was then controlled using MMseqs2 with the UniqueProt procedure to reduce homologous overlap, limiting the sequence identity within the test set and between the training and test sets to 20% while preserving diversity in the training data (Olenyi et al., 2022). The final curated dataset contained 4,089 acidophilic proteins and 1,654 non-acidophilic proteins, which were subsequently split into training and testing subsets (80/20), with a training dataset consisting of 3271 positive samples and 1323 negative proteins, and an independent test set consisting of 818 positive samples and 331 negative samples for an unbiased evaluation.

### 2.2 Protein Language Model

Protein language models (PLMs) extend the self-supervised learning paradigm of Transformer-based language modeling to biological sequences by treating amino-acid strings as a “protein language” and training on large unlabeled corpora to learn contextual representations. This framework produces embeddings that work well across many protein-related downstream tasks and offers a strong alternative to traditional methods built on hand-crafted features and shallow models. (Rives et al., 2021, Devlin et al., 2018).

This study adopts ESM C-600M, an upgraded model in the ESM family, which aims to better model the fundamental biological characteristics of proteins. Through larger training datasets and an increased computational scale, it delivers marked performance gains over ESM2 (ESM Team, 2024). ESM C embeddings capture higher-order contextual and evolutionary patterns, including motif–context interactions, non-local residue couplings, and family-level signals that emerge from shared evolutionary constraints. In practice, ESM C provides strong and biologically informed initialization for supervised classification. Rather than learning protein representations from scratch on a relatively small labeled acidophilic dataset, the model leverages pre-trained attention and feed-forward parameters that already encode general protein regularities.

### 2.3 Parameter-Efficient Fine-Tuning

To adapt the PLM to this specific classification mission under limited computational budgets, we adopted parameter-efficient fine-tuning (PEFT), a family of fine-tuning approaches that enables task adaptation by training only a small set of additional parameters while keeping most of the pretrained weights frozen. These approaches are particularly effective for protein sequence modeling, where Transformer backbones are large and input lengths can be long, making end-to-end updating prohibitively expensive in many practical settings. In this study, we benchmarked six PEFT methods. LoRA completes task-specific adaptation by injecting trainable low-rank update matrices into selected linear operators, parameterizing the weight increment as $\Delta W = BA$ with $rank \ll min(d_{in}, d_{out})$, so the effective transformation becomes $W' = W + \Delta W$ while only optimizing B and A (Hu et al., 2021). MiSS, $C^3A$, OFT, RandLoRA, and SHiRA are five other LoRA-based PEFT methods, as listed in **Table 1**. We replaced the key linear layers of high dimension with the PEFT framework, including the query, key, value projections, attention output projection, and adapted major dense transformations in the feed-forward sublayers as well as the corresponding normalization-associated projections. After attaching the PEFT modules to these layers, the backbone parameters remained largely frozen, with the proportion of trainable parameters controlled within 3% for all six methods.

**Table 1** Comparison of PEFT methods showing their forward mappings and initialization schemes.

| Method | Forward | Initialization |
|---|---|---|
| LoRA | $y = W_0 x + BAx$ | $A \sim N(0, \sigma^2), B = 0$ |
| MiSS | $y = W_0 x + expand(D)x$ | $D = 0$ |
| $C^3A$ | $y = W_0 x + \Delta W x$<br>$\Delta W = C(\Delta w)$ | $\Delta W = 0$ |
| OFT | $y = (W_0 R)x$<br>$R = (I + Q)(I - Q)^{-1}$ | $Q = 0 \Rightarrow R = I$ |
| RandLoRA | $y = W_0 x + \Delta W x$<br>$\Delta W = \sum_j B_j \Lambda_j A \Gamma_j$ | $A, B_j$ $random$, $\Lambda_j$, $\Gamma_j$ $trainable$ |
| SHiRA | $y = (W_0 + M \odot \Delta W)\, x$ | $\Delta W = 0, M$: *fixed mask* |

### 2.4 DeepSeekMoE

In our setting, although PEFT substantially reduces the number of trainable parameters in the backbone, the original classification head remains a major source of parameter overhead. In particular, the classification head couldn't be replaced and included a large dense mapping with input dimension 2304 and output dimension 4608, which contributed on the order of $2304 \times 4608 \approx 1.06 \times 10^7$parameters. This scale was comparable to the total number of trainable parameters introduced by PEFT (e.g., LoRA in our configuration is roughly $1.2 \times 10^7$), meaning that even if the backbone is adapted efficiently, the full classifier can dominate the trainable budget. To reduce this burden, we replaced the original dense classifier with a DeepSeekMoE-based head.

The DeepseekMoE head adopts a sparse expert-routing scheme rather than a single dense transformation. Given the pooled representation $h$, a lightweight gate assigns routing weights to multiple experts and activates only the top-$k$ experts for each sample. The output can be written as

$$y = \sum_{i \in \text{Top-}k} \alpha_i \, E_i(h), \tag{1}$$

where $E_i(\cdot)$ denotes the $i$-th expert and $\alpha_i$ is the corresponding routing weight. In our implementation, we used three routed experts and selected two for each input. Each expert was modeled as a lightweight two-layer MLP, allowing the head to preserve the expressive capacity without introducing a large number of additional parameters.

Besides the routed experts, the head also includes a shared expert branch that is always active. This branch provides a stable transformation path that is independent of routing decisions and complements the sparse expert outputs. To improve the training stability, we further introduced an auxiliary load-balancing objective so that expert usage would not become overly concentrated on only one or two experts. This is particularly useful in imbalanced classification settings, where unstable routing may weaken the benefit of the expert structure.

This design retains a stable capacity independent of routing decisions while still enabling conditional computation through the routed experts. As a result, the number of trainable parameters significantly decreased, as shown in **Table 2**, leading to less allocated GPU memory and training time.

**Table 2** The number and proportion of trainable parameters, average epoch time and allocated GPU memory of six MI-PEFT architectures.

| | Trainable Parameters | Proportion | Average Epoch Time | GPU memory allocated |
|---|---|---|---|---|
| LoRA | 3212552 | 0.56% | 78.3s | 27811.95MB |
| MiSS | 2549000 | 0.44% | 76.8s | 25121.87MB |
| $C^3A$ | 2549000 | 0.44% | 101.6s | 25792.23MB |
| OFT | 2507528 | 0.43% | 76.1s | 25948.23MB |
| RandLoRA | 4290824 | 0.74% | 102.7s | 26186.20MB |
| ShiRA | 5203208 | 0.90% | 78.8s | 26775.40MB |

## 2.5 Temperature Scaling

To prevent the model from becoming over confident, we applied Temperature Scaling, a straightforward and effective post-hoc calibration technique (Guo et al., 2017). While our model outputs logits $z$ for classification, the softmax probabilities derived from them are often poorly calibrated, meaning that they do not accurately reflect the true likelihood of a prediction. Temperature Scaling addresses this by introducing a single, positive scaling parameter T (temperature) to soften or sharpen the output distribution. The calibrated probability for class i is computed as:

$$q_i = \frac{exp(z_i/T)}{\sum_j exp(z_j/T)} \tag{2}$$

where $T > 1$ produces a softer, more conservative probability distribution, and $0 < T < 1$ yields a more confident peaking distribution. The optimal temperature T is learned by minimizing the negative log likelihood (NLL) on a held-out validation set while keeping all other model parameters frozen. This optimization was performed efficiently using the L-BFGS algorithm (Liu & Nocedal, 1989).

In our framework, this process was integrated into each fold's evaluation pipeline. After completing the training and selecting the best model checkpoint based on the validation loss, we reloaded the optimal weights and performed a forward pass on the validation set to obtain the raw logits. These logits, together with the true labels, were used to fit the fold-specific temperature parameter $T$. Subsequently, all predictions of the independent test set were calibrated using this learned $T$ before the final metric computation and probability-based analyses.

## 2.6 Performance Evaluation

We employed a suite of standard metrics, visualization techniques, and statistical analyses to comprehensively evaluate the performance of our proposed model for acidophilic protein classification. The evaluation is conducted during 10-fold cross validation and eventually on an independent test set, which was completely sequestered during all phases of model training and hyperparameter optimization, thereby providing an unbiased estimation of the model's ability to generalize to unseen data and mitigating the risk of overfitting (Kohavi, 1995). We presented a confusion matrix for each model to offer a detailed breakdown of the model's predictions versus true labels (Stehman, 1997). This matrix provides direct insight into the number of specific types of predictions made by the classifier, including True Positives (TP), True Negatives (TN), False Positives (FP), and False Negatives (FN). Visual inspection reveals whether misclassifications are symmetric or if the model exhibits a bias toward one class. Based on the confusion matrix, we calculated six standard performance metrics: Accuracy (**Eq. 3**), Sensitivity (**Eq. 4**), Specificity (**Eq. 5**), Matthews Correlation Coefficient (**Eq. 6**), F1-Score (**Eq. 7**), and Area Under the Receive Operating Characteristics Curve (AUC), which encapsulates the probability that the model ranks a random positive instance higher than a random negative instance. Accuracy (**Eq. 3**) represents the overall proportion of correct predictions.

$$Acc = \frac{(TP + TN)}{(TP + TN + FP + FN)} \tag{3}$$

$$Sn = \frac{TP}{(TP + FN)} \tag{4}$$

$$Sp = \frac{TN}{(TN + FP)} \tag{5}$$

$$MCC = \frac{(TN \times TP - FP \times FN)}{\sqrt{(TP + FP)(TP + FN)(TN + FP)(TN + FN)}} \tag{6}$$

$$F1 - Score = 2 \times \frac{\left(\frac{TP}{TP + FN}\right) \times \left(\frac{TP}{TP + FP}\right)}{\left(\frac{TP}{TP + FN}\right) + \left(\frac{TP}{TP + FP}\right)} \tag{7}$$

To examine the learned representations, we extracted embeddings from the final hidden layer of the Transformer encoder and projected them into two dimensions using UMAP (McInnes et al., 2018). This visualization was used to assess class separation and potential ambiguous regions in the feature space. We further analyzed the density distribution of prediction confidence, defined as the softmax probability of the predicted class, to evaluate calibration and confidence reliability (Guo et al., 2017).

## 2.7 Experimental Setup

The experimental framework was implemented in Python using PyTorch and the Hugging Face transformer library. All model training and evaluation were conducted on a single NVIDIA GeForce RTX 5090 GPU (24GB VRAM). To accelerate training and maximize memory efficiency, computations were performed using mixed-precision training via PyTorch Automatic Mixed Precision (AMP).

Model optimization was performed using the AdamW optimizer, which incorporates decoupled weight decay to improve regularization and convergence (Loshchilov & Hutter, 2019). Considering the overall length distribution of proteins in the dataset, we truncated the maximum length to 512 tokens. A base learning rate of $5 \times 10^{-5}$ was employed ($5 \times 10^{-4}$ for $C^3A$), with a weight decay factor of 0.01 to mitigate overfitting. The learning rate schedule followed a linear decay with warm-up: for each training fold, the first 10% of the total optimization steps were allocated to a warm-up phase where the learning rate increased linearly from zero to the peak value, after which it decreased linearly to zero over the remaining steps. This strategy helps stabilize training in the initial phase. The effective batch size was set to 32, which was achieved through gradient accumulation, allowing the simulation of a larger batch size within the GPU memory constraints. The specific key configurations for each PEFT method are listed in **Table 3**.

The model was trained for a maximum of 10 epochs per cross-validation. To prevent overfitting and determine the optimal stopping point, an early stopping mechanism was implemented with a patience of 3 epochs. This mechanism continuously monitored the cross-entropy loss on the validation set; if no improvement was observed for three consecutive epochs, training would be halted, and the model weights from the epoch with the lowest validation loss were restored for final evaluation of the independent test set.

**Table 3** Key hyperparameter configurations for each PEFT method used in this study.

| **Methods** | LoRA | MiSS | $C^3A$ | OFT | RandLoRA | ShiRA |
|---|---|---|---|---|---|---|
| **Key Parameters** | *r = 8*<br>*α = 16*<br>*dropout = 0.01* | *r = 8*<br>*dropout = 0.01* | *block size = 128* | *block size = 32* | *r = 32*<br>*α = 128*<br>*Seed = 0* | *r = 16*<br>*mask = random* |

2.8 Baseline Models

To provide a meaningful comparison of the proposed PLM-based method, we evaluated a set of conventional machine learning and deep learning baseline models using standard input representations and training protocols. These baselines were designed to assess how commonly used classifiers can capture discriminative signals for acidophilic protein prediction without task-specific fine-tuning of large pretrained models. Specifically, we considered six classical machine learning methods: Logistic Regression (Cox, 1958), Support Vector Machine (Cortes & Vapnik, 1995), Random Forest (Breiman, 2001), Extreme Gradient Boosting (Chen & Guestrin, 2016), LightGBM (Ke et al., 2017), and SVM with a radial basis function kernel. Additionally, we adopted three deep learning baselines: a feedforward deep neural network (Rumelhart et al., 1986), a unidirectional LSTM, and a bidirectional LSTM (Hochreiter & Schmidhuber, 1997). All baseline models were evaluated under consistent data splits and evaluation criteria, enabling direct comparison with the proposed MI-PEFT($C^3A$) and SOTA.

## 3 Results and Discussion

3.1 The dataset distribution demonstrates class imbalance

We first examined the distributional characteristics of the dataset to explore the inherent differences between positive and negative samples. The dataset is characterized by a significant class imbalance with acidophilic proteins substantially outnumbering negative samples by roughly three times in both the training and testing sets because there're fewer and more scattered non-acidophilic proteins retrievable from the dataset. This imbalance poses a significant challenge for classification, as naive optimization strategies may favor the majority class to maximize the overall accuracy at the expense of specificity.

**Fig. 1A** shows that the majority of proteins consist of less than 200 amino acids. Positive samples dominate this region, whereas negative samples are more dispersed across longer sequence ranges. Also, the amino acid composition profiles of positive and negative samples shows that both classes share similar overall trends. There's a slight difference that acidophilic proteins show relatively higher proportions of acidic residues (such as aspartic acid and glutamic acid), whereas negative samples exhibit comparatively increased frequencies of certain hydrophobic and neutral residues. However, the observed compositional shifts are moderate with substantial overlap between two classes, indicating that acidophilicity is not determined simply by the proportion of amino acids. The dataset is input to our framework in **Fig. 1B.** An overview of the performance is shown in **Fig. 1C**, which will be discussed in detail. The complete dataset and framework are available at https://github.com/Duncan-SHH/MI-PEFT-for-Acidophilic-Protein-Classification/tree/main for reference and for further exploration.

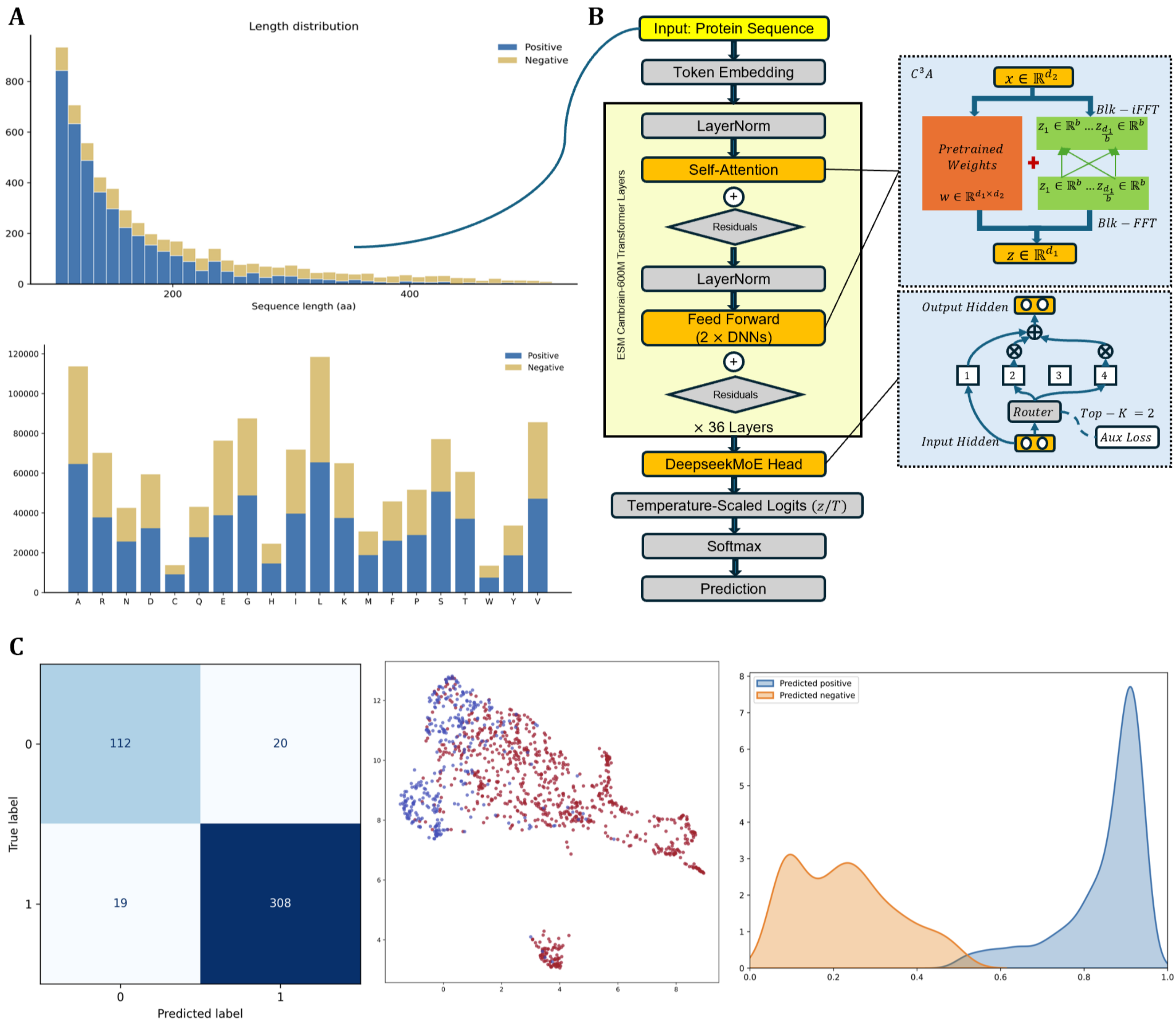

**Figure 1** (A) Length distribution and amino acids composition for positive and negative samples. (B) Overall architecture of the model. Protein sequences are first encoded by the ESM C-600M backbone, followed by a DeepSeekMoE classification head for efficient parameter utilization. Temperature scaling is applied to logits during evaluation to improve confidence calibration before final prediction. (C) Overview of MI-PEFT($C^3$A) performance.

### 3.3 $C^3A$ performs the best on both ten-fold cross-validation and an independent test set

We compared six PEFT strategies under two evaluation protocols: ten-fold cross-validation on the training data and testing on an independent held-out set. According to **Fig. 2**, the results indicate that all six methods deliver highly comparable predictive performances. Under ten-fold cross-validation, the accuracy remained tightly clustered in the range of approximately 0.90–0.91. SN was consistently high for all methods (approximately 0.935–0.940), while SP was lower and more variable than SN, indicating that discriminating non-acidophilic proteins is relatively difficult due to the imbalanced dataset, which increases the risk of false positives. According to the MCC (approximately 0.76–0.79) and F1-score (approximately 0.93–0.94), the overall decision boundary is slightly skewed but still well balanced between precision and recall. AUC values were uniformly strong (approximately 0.95–0.96), implying that all six methods can produce well-separated score distributions even when threshold-dependent metrics such as SP fluctuate.

The independent test set evaluation largely reproduced the trend observed in cross-validation with stronger stability. The margin error in independent test set evaluation was smaller for all metrics and all methods because of consistency. Accuracy on the independent set was again tightly grouped around 0.91–0.92, and SN remained near 0.94–0.95 across all methods. SP continued to be the metric with the largest spread, with most methods concentrated near 0.84–0.85. MCC and F1-score followed the same stable pattern as cross-validation, while AUC values remained uniformly high and close across methods, reinforcing that none of the six approaches suffered a meaningful degradation in ranking performance.

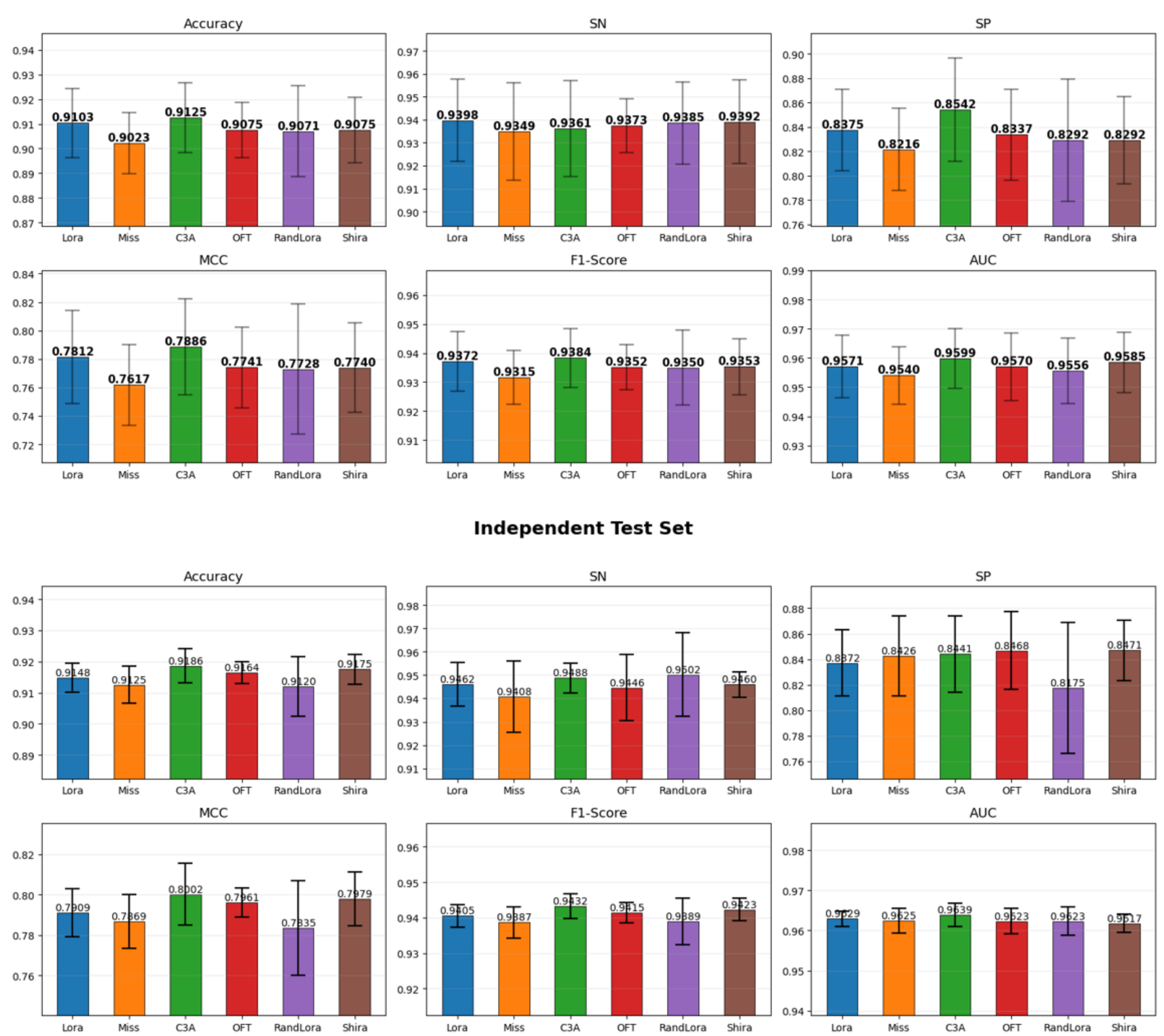

**Figure 2** Comparison of six evaluation metrics (Accuracy, SN, SP, MCC, F1-score, and AUC) across six methods in 10-fold cross-validation and on the independent test set.

The results indicate that specificity is the principal limitation in model performance, which can be primarily attributed to the class imbalance. Among six PEFT methods, although the differences in accuracy were not dramatic, $C^3A$ especially stood out in SP and MCC, from which we can conclude that it's the most adaptable one to the class-imbalanced setting. Besides, $C^3A$ required relatively small proportion of trainable parameters and low GPU memory. Although it takes up more training time, balancing both effectiveness and efficiency, $C^3A$ was selected as the best-performing method and we're using its performance for further analysis and comparison in this study.

### 3.4 MI-PEFT($C^3A$) Outperforms Machine Learning and Deep Learning Baselines and the ACE Model Under the Same PLM

To further assess the effectiveness of our approach, we further compared the MI-PEFT($C^3A$) against a broad set of classical machine learning models and deep learning baselines. We additionally evaluated the ACE model by fine-tuning the same ESM C backbone and assessing its performance on an identical independent test set to control a single variable. Under this controlled setting, MI-PEFT($C^3A$) still demonstrates consistent advantages across most metrics, especially in terms of specificity. As shown in **Fig. 3**, our method achieves the best or near-best scores across all the six evaluation metrics. In particular, classical machine learning models that rely on simple and fixed handcrafted features exhibit pronounced deficiencies in specificity and MCC, indicating a limited ability to cope with the class imbalance of the dataset. These models tend to favor predicting samples as positive to inflate overall accuracy, which leads to a substantial increase in false positives and consequently poor specificity and an imbalanced trade-off between sensitivity and specificity.

Deep learning baselines, including DNN, LSTM, and BiLSTM, show improved sensitivity and F1-score compared with classical machine learning methods, but still demonstrated the same shortcoming. Despite additional local hyperparameter tuning, deep learning models tend to overfit training data for higher accuracy. In contrast, models built on PLM achieved substantially improved and stable performance. By leveraging self-supervised pretraining on large-scale protein sequence corpora, PLMs can capture general biochemical, structural, and evolutionary regularities that are difficult to encode through handcrafted features or learn reliably from limited labeled data alone. This advantage is particularly evident under class-imbalanced settings, where models without previous knowledge are trained merely on given dataset.

Compared to the ACE framework, MI-PEFT($C^3A$) exhibits an explicitly better balance between sensitivity and specificity. While both approaches benefit from pretrained protein representations, MI-PEFT($C^3A$) achieves improved discrimination of negative samples. ACE fine-tuning allows a broader modification of the representation space, which can inadvertently shift the decision boundary toward the majority (positive) class, increasing false positives and degrading specificity. In contrast, out method constrains task-specific adaptation to low-rank transformations applied to selected projection layers, limiting the extent to which pretrained representations are globally distorted. This restricted adaptation prevents excessive expansion of the positive decision region. As a result, MI-PEFT($C^3A$) maintained high sensitivity while preserving better discrimination of negative samples, leading to a more balanced trade-off between sensitivity and specificity.

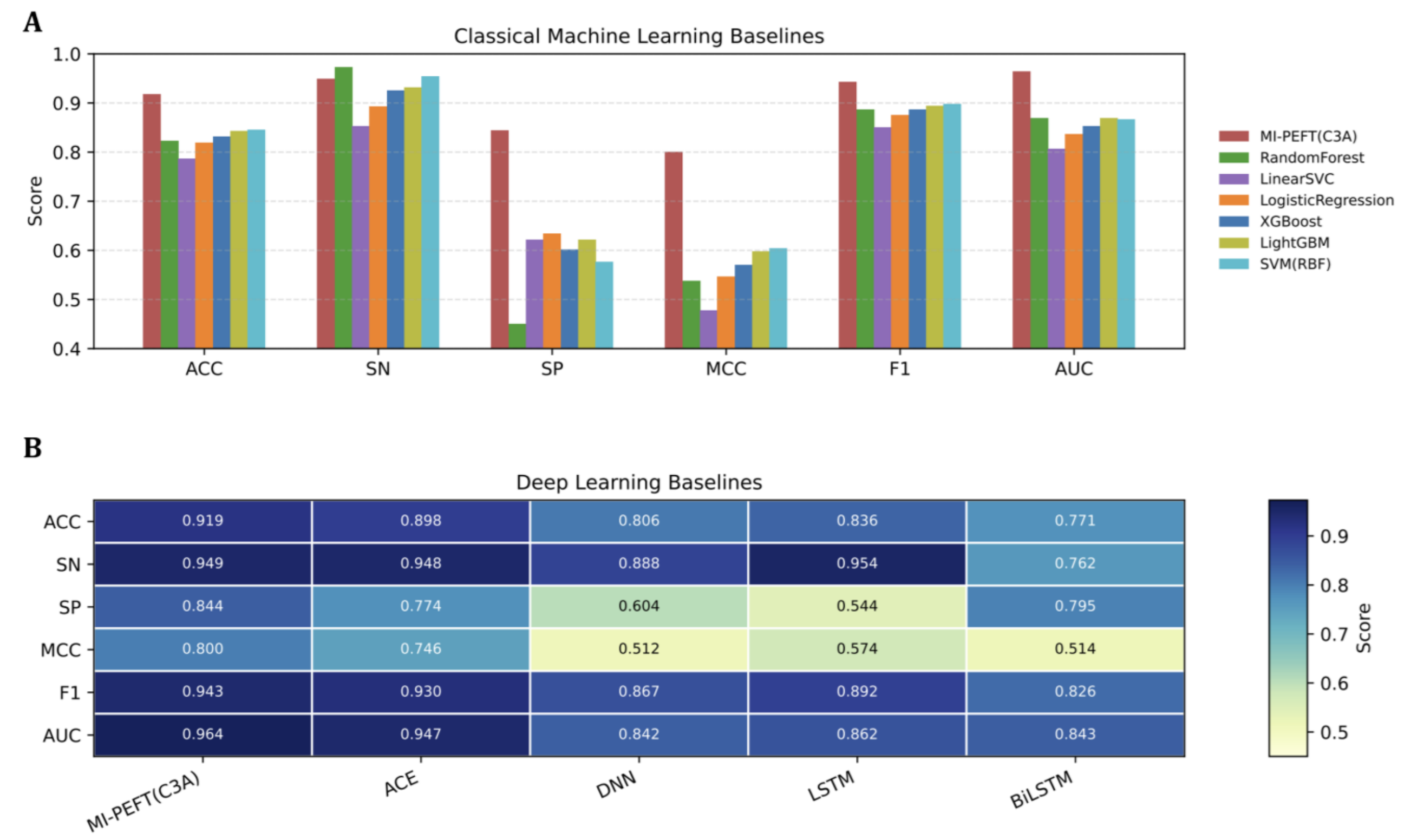

**Figure 3** Performance comparison of MI-PEFT($C^3$A) with baseline models on the independent test set. (A) Comparison with ML baseline models. (B) Comparison with DL baseline models.

## 3.5 Ablation studies of MI-PEFT show the effectiveness of components

Two ablation studies were conducted to demonstrate the effectiveness of the selected components. First, to determine the best substitute for the classification head, we made a comparison between the DeepSeekMoE, Vanilla MoE and the MLP baseline.

**Table 4** Comparison of head architectures and key hyperparameters for DeepSeekMoE, Vanilla MoE, and the MLP in our settings.

| **Setting** | DeepSeekMoE | Vanilla MoE | MLP |
|---|---|---|---|
| **Expert Structure** | $Linear \rightarrow ReLU \rightarrow Linear$ | $Linear \rightarrow ReLU \rightarrow Dropout \rightarrow Linear$ | $MLPHead$ |
| **Expert Branch** | $Shared\ Expert + Routed\ Experts$ | $Routed\ Experts\ Only$ | / |
| **Expert Hidden Dimension** | 128 | 196 | / |
| **MLP Hidden Dimension** | / | / | 512 |
| **Head Parameters** | 1221896 | 1363434 | 1181186 |

Compared to DeepSeekMoE, Vanilla MoE relies solely on routed experts. Besides, DeepSeekMoE contains an auxiliary load balancing loss. We controlled the hidden dimensions as shown in **Table 4** so that their head parameter numbers are at the same level.

In terms of performance **(Fig. 4A)**, DeepSeekMoE achieved the highest score for all the six metrics. The other two baselines lag especially in sensitivity and MCC, suggesting that they miss more positives and get more influenced by the class-imbalanced dataset. One plausible explanation is that in a standard MoE head, class imbalance can bias the routing process because most updates come from majority class samples. This can cause the gate to send many inputs to a small number of frequently used experts, while others remain under trained. The shared expert pathway mitigates this risk by providing a representation that is always trained by all samples, so minority signals are not entirely dependent on routing. Simultaneously, the auxiliary load balancing loss ensures more even expert usage, reducing the chance that learning concentrates on only a few experts.

We further compared the performances of different ESM family PLMs under MI-PEFT($C^3$A), including ESM C-600M, ESM C-300M, ESM2-35M, ESM2-150M, ESM2-650M, and ESM2-3B. As shown in **Fig. 4B**, ESM C-600M yielded the strongest performance in all metrics, with a particularly noticeable edge in the MCC. ESM C 300M ranks second and remains competitive on all metrics, outperforming the older ESM family models despite using substantially fewer parameters.

Noticeably, ESM2-3B did not perform better than ESM2-650M or ESM2-150M despite its large-scale parameter. Only ESM2-35M demonstrated evident drawbacks. These results suggest that in the PEFT setting, a parameter scale is necessary but isn't sufficient for outstanding downstream task performance. Also, the evident benefit of ESM C is likely because it's trained on a larger dataset and has a stronger emphasis on generalized features. This feature allows ESM C to perform more stable under class-imbalanced settings because it's less dominated by sequence patterns that appear at high frequencies.

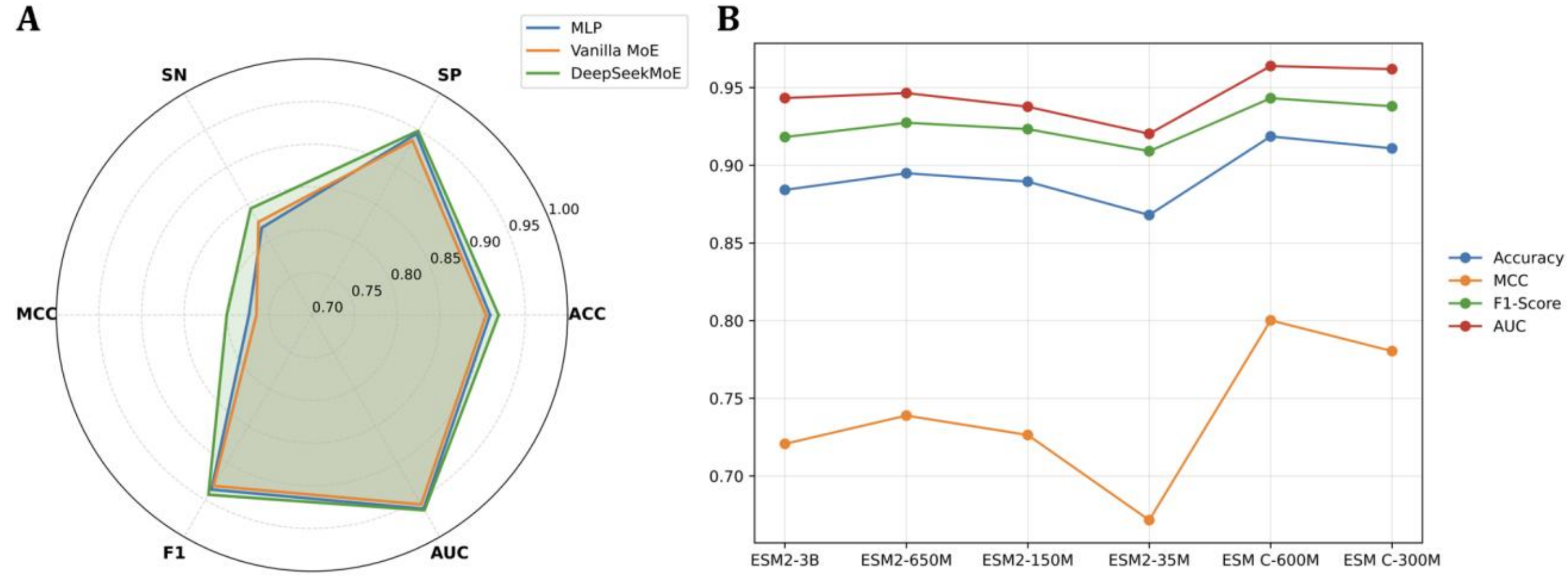

**Figure 4** (A) Radar chart comparing the performance of MLP, Vanilla MoE, and DeepSeekMoE across six evaluation metrics. (B) Performance evaluation for different ESM family models under MI-PEFT($C^3$A).

3.7 Advantage of MI-PEFT($C^3$A) Over the SOTA is Further Supported by UMAP and Prediction Confidence Distributions

In the UMAP projections derived from the final hidden layer (**Fig. 5A**), MI-PEFT($C^3A$) produced more compact and class-consistent clusters, with reduced overlap between acidophilic and non-acidophilic proteins. In contrast, the ACE model exhibits a more diffuse embedding structure, where samples from different classes partially intermingle, suggesting that the linear adaptation of frozen PLM embeddings is insufficient for fully separating subtle sequence-level signals.

The structural difference was mirrored by the confidence density plots (**Fig. 5B**). Under ACE fine-tuning, the predicted probabilities for positive samples are closer to one, while negative samples are sharply skewed toward zero. MI-PEFT($C^3A$), by comparison, shows broader and flatter confidence distributions for both classes through temperature scaling. In practice, this clearerer separation in prediction confidence provides more informative guidance for selecting candidate proteins for experimental validation, which is typically time-consuming and labor-intensive.

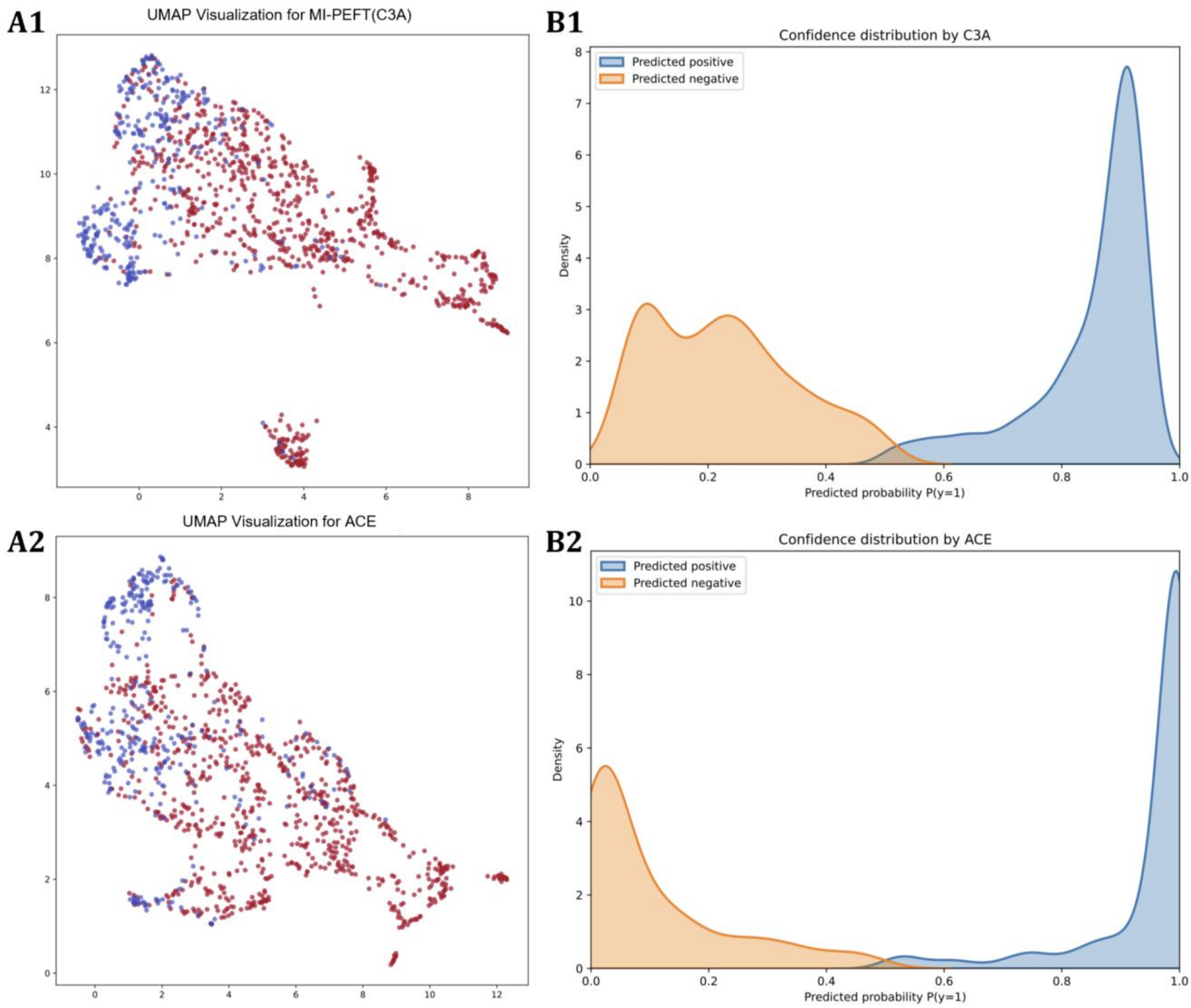

**Figure 5** Advanced visualization techniques for comparison with SOTA. (A) UMAP of learned representations, with red dots indicating positive samples and blue dots indicating negative samples. (B) Density distribution of prediction confidence.

## Conclusion

In this study, we propose MI-PEFT, a mixture-of-experts integrated parameter-efficient fine-tuning framework for the identification of acidophilic proteins. By leveraging the strong representation capability of protein language models and constraining parameter updates through PEFT strategies, the framework effectively balances predictive performance and computational efficiency. Systematic comparisons across six PEFT methods demonstrate that $C^3A$ achieves the most favorable overall performance, particularly in specificity and MCC, which are critical in the class imbalanced setting of this task.

In addition, the integration of a DeepSeekMoE classification head further enhances the capacity of the model to capture heterogeneous sequence patterns, while temperature scaling improves the reliability of the prediction confidence. Overall, this study highlights the effectiveness of combining pretrained protein representations with constrained adaptation and expert routing mechanisms for challenging biological sequence classification problems. The proposed framework not only provides a practical tool for accurate identification of acidophilic proteins but also offers a general paradigm for applying parameter efficient adaptation strategies to protein language models in scenarios with limited and imbalanced data.